\documentclass[aps,prb,twocolumn,showpacs,superscriptaddress,groupedaddress]{revtex4}%aps,prb,preprint]{revtex4}
\usepackage{hyperref}
\usepackage[T1]{fontenc}
\usepackage[utf8]{inputenc}
\usepackage{graphicx}  % needed for figures
\usepackage{dcolumn}   % needed for some tables
\usepackage{bm}        % for math
\usepackage{amssymb}   % for math
\usepackage{amsfonts}
\usepackage{epsfig}
\usepackage{mathrsfs}
\usepackage[fleqn,tbtags]{amsmath}
\usepackage{mathtools}
\usepackage{enumerate}
\usepackage{color}
\usepackage{siunitx}

\usepackage{xcolor}
\newcommand{\ve}[1]{\mathbf{#1}}
\newcommand{\vs}[1]{\boldsymbol{#1}}

\newcommand{\ex}[1]{\langle #1 \rangle}
\newcommand{\ann}[2]{#1^{\phantom{\dagger}}_{#2}}
\newcommand{\cre}[2]{#1^\dagger_{#2}}

\newcommand{\im}{\text{Im}}

\begin{document}

\title{Theory of Curie temperature enhancement in electron-doped EuO}

\author{Tobias Stollenwerk}

\affiliation{Physikalisches Institut and Bethe Center for Theoretical Physics,
  Universit\"at Bonn, Nussallee 12, 53115 Bonn, Germany}
\affiliation{German Aerospace Center, Linder H\"ohe, 51147 Cologne, Germany}
\author{Johann Kroha}
\email[Email: ]{kroha@physik.uni-bonn.de}
\affiliation{Physikalisches Institut and Bethe Center for Theoretical Physics,
  Universit\"at Bonn, Nussallee 12, 53115 Bonn, Germany}
\affiliation{Center for Correlated Matter, Zhejiang University, 
Hangzhou, Zhejiang 310058, China}

%\date{\today}
\received{10 September 2015; revised manuscript received 20 October 2015;
published 17 November 2015}

\begin{abstract}
We present a comparative, theoretical study of the doping dependence 
of the critical temperature $T_C$ of the ferromagnetic insulator-metal 
transitions in Gd-doped and O-deficient EuO, respectively. 
The strong $T_C$ enhancement in Eu$_{1-x}$Gd$_x$O is due to 
Kondo-like spin fluctuations on the Gd sites, which are absent 
in EuO$_{1-x}$. Moreover, we find that the $T_C$ saturation in 
Eu$_{1-x}$Gd$_x$O for large $x$ is due to a reduced activation 
of dopant electrons into the conduction band, in agreement 
with experiments, rather than antiferromagnetic long-range 
contributions of the RKKY interaction. The results shed light 
on possibilities for further increasing $T_C$.
\end{abstract}

\maketitle

%%%%%%%%%%%%%%%%%%%%%%%%%%%%%%%%%%%%%%%%%%%%%%%%%%%%%%%%%%%%%%%
%%%%%%%%%%%%%%%%%%%%%%%%%%%%%%%%%%%%%%%%%%%%%%%%%%%%%%%%%%%%%%%
%%%%%%%%%%%%%%%%%%%%%%%%%%%%%%%%%%%%%%%%%%%%%%%%%%%%%%%%%%%%%%%
%  Introduction
%%%%%%%%%%%%%%%%%%%%%%%%%%%%%%%%%%%%%%%%%%%%%%%%%%%%%%%%%%%%%%%
%%%%%%%%%%%%%%%%%%%%%%%%%%%%%%%%%%%%%%%%%%%%%%%%%%%%%%%%%%%%%%%
%%%%%%%%%%%%%%%%%%%%%%%%%%%%%%%%%%%%%%%%%%%%%%%%%%%%%%%%%%%%%%%
\section{Introduction}
\label{sec:introduction}
The demand for ever increasing speed and integrability of magnetic 
information storage devices as well as other spintronics applications calls 
for materials that are capable of transforming electronic or optical 
signals efficiently into magnetization and vice versa.  
Electron-doped europium monoxide (EuO) is a promising candidate 
for this purpose, as it undergoes a simultaneous ferromagnetic (FM) and 
insulator-to-semimetal transition [\onlinecite{Petrich71}],
exhibiting an outstanding magneto-electric response, including the strongest 
colossal magnetoresistance effect known [\onlinecite{Reed_a73,Reed_b73}], 
magneto-optical effects [\onlinecite{Ahn67,Schoenes74,Matsubara10,Matsubara15}], 
and a spin polarization of the conduction band of nearly 
100~\% in the FM state [\onlinecite{Steeneken02,Schmehl07}]. 
Improved sample fabrication techniques [\onlinecite{Foerster11,Klinkhammer13}] 
and europium monoxide's epitaxial integrability into Si [\onlinecite{Schmehl07}] 
and GaAs [\onlinecite{Awschalom10}] structures have renewed and intensified 
the interest in this material during the past few years.

Stoichiometric EuO is a wide band gap  
semiconductor with rocksalt structure. The local magnetic moments 
of $m=7/2~\mu_B$ situated in the Eu $4f$ shell constitute a 
prototype Heisenberg ferromagnet, ordering ferromagnetically at the 
Curie temperature of $T_C=69$~K. Their interaction is 
mediated by virtual excitations (hybridization) of the tightly bound Eu $4f$ 
electrons into the spatially more extended Eu $5d$ orbitals and an 
exchange between the latter [\onlinecite{Kasuya72}]. 
Upon electron doping the FM transition is accompanied by a 
simultaneous insulator-to-semimetal transition with a 
resistivity drop by 8 to 13 orders of magnitude
[\onlinecite{Oliver2,Penney72,Steeneken02}]. Raising the transition 
temperature significantly is one of the major challenges involved 
in bringing the extraordinary properties of doped EuO to technological use. 

Since early on, the general trend of $T_C$ being increased by electron-doping 
has been associated with the formation of magnetic polarons
[\onlinecite{Mauger77,Mauger86}], i.e., conduction electrons dragging along 
a magnetic polarization cloud of local Eu $4f$ moments. However, the experiments
reveal more complex behavior. 
Gadolinium doping replaces Eu by Gd atoms, leaving the 
lattice of magnetic $4f$ moments intact and donating one 
additional electron per Gd atom from the Gd $5d$ shell. In the doped 
material, Eu$_{1-x}$Gd$_{x}$O, $T_C$ increases to values 
between $120$~K and $170$~K for Gd concentrations in the range of 
$x=0.04,\dots , 0.08$
[\onlinecite{Schoenes74,Matsumoto04,Ott06,Mairoser10,Altendorf12,Mairoser13}], 
depending on the sample quality, strain, and measurement 
conditions [\onlinecite{Mairoser13}]. Invariably, $T_C$ saturates for higher $x$.
Oxygen defects introduce two electrons per O defect, 
but only a weak $T_C$ increase has been observed in bulk, 
Eu-rich EuO$_{1-x}$
[\onlinecite{Oliver2,Reed_a73,Reed_b73,Penney72,Altendorf11,Barbagallo11}]. The 
$T_C$ increase reported in Ref.~[\onlinecite{Barbagallo10}] for EuO$_{1-x}$ 
may presumably be attributed [\onlinecite{Mairoser13}] to the presence of a large 
external magnetic field inherently necessary for the SQUID measurement 
technique used. 

The magnetic polaron theory alone cannot account for the $T_C$ saturation 
at high Gd concentration nor for the fact that O defects essentially do not 
raise $T_C$, even though they introduce twice as many carriers per impurity 
as Gd doping. 
It has been proposed [\onlinecite{Kogan11,Schwingenschloegl15}] 
that the $T_C$ saturation might be
understood in that, for increasing conduction-band filling, the oscillatory RKKY
interaction [\onlinecite{RKKY1,RKKY2,RKKY3}] acquires increasingly antiferromagnetic 
(AF) contributions due to the decreasing RKKY wavelength. This requires, 
however, unrealistically high band filling. On the other hand, Hall resistivity
measurements indicate [\onlinecite{Mairoser10}] that the density of mobile charge 
carriers activated into the conduction band saturates in line with the 
$T_C$ saturation, providing a phenomenological reason for the 
limited $T_C$ increase in Eu$_{1-x}$Gd$_x$O.
In theoretical calculations, treating the O vacancies in a static 
approximation, Sinjukow and Nolting [\onlinecite{Sinjukow04}] found 
no increase of $T_C$ in EuO$_{1-x}$ for an appropriate choice of system 
parameters. More sophisticated resummation techniques for the Gd impurity 
potential in Eu$_{1-x}$Gd$_x$O were able to correctly describe a shallow 
maximum of $T_C$ as a function of Gd concentration [\onlinecite{Takahashi12}], 
but not the saturation of the mobile charge carrier density [\onlinecite{Mairoser10}.  Recent {\it ab initio} calculations
[\onlinecite{Belashchenko13,Schwingenschloegl15}] provided more quantitative results
on the coupling parameters and spectral densities, but did not lead to 
a consistent understanding of all the experimental facts described above. 
Taking the local spin fluctuations on Gd impurities into
account, Arnold and Kroha [\onlinecite{Arnold08}] could explain details of the 
magnetization behavior, the simultaneity of the insulator-semimetal 
transition and the $T_C$ increase in Eu$_{1-x}$Gd$_x$O.

In this paper we report a comprehensive, theoretical study of the 
$T_C$ enhancement in electron-doped EuO, extending the model of  
Ref.~[\onlinecite{Arnold08}]. Gd impurities as well as O defects are treated 
dynamically as Anderson impurities hybridizing with the conduction band,
however with strong or moderate on-site repulsion, respectively,
ensuring the single or double occupancy of the Gd impurity or O vacancy 
orbitals. 
The direct comparison of the two cases shows that indeed the $T_C$ increase 
with Gd doping is caused by the Kondo-like spin fluctuations on the Gd sites 
and the concatenated accumulation of impurity spectral weight as well as conduction electron 
spectral weight at the chemical potential. This dynamical many-body effect 
drives the metallic transition, which, in turn, enhances the polaronic 
FM coupling between the Eu moments. By contrast, on O vacancies 
the two defect electrons form a spin-singlet, and local spin fluctuations 
are absent, leading only to a moderate $T_C$ enhancement due to a weak 
population of the conduction band. Moreover, the theory explains that in 
Eu$_{1-x}$Gd$_x$O the activation of charge carriers into the conduction band 
decreases with increasing doping concentration, in agreement with 
experiments [\onlinecite{Mairoser10}], leading to the saturation of $T_C$. 
Including explicitly the RKKY interaction in our 
theory, we find that for all relevant temperatures $T$ and 
doping concentrations (band fillings) its long-distance AF contributions 
are negligible.
  
The paper is organized as follows. 
In Sec. II, we give a detailed justification of our model 
for Eu$_{1-x}$Gd$_x$O and EuO$_{1-x}$ and describe the theory 
for its evaluation.
The results are shown and discussed in Sec. III. 
We conclude in Sec. IV with a suggestion for a possible pathway 
to further enhance the transition temperature in 
electron-doped EuO.
%The appendix contains a general derivation of RKKY interaction
%and the equations of the non-crossing approximation, adapted
%for the description of a finite density of Gd impurities. 
 
%%%%%%%%%%%%%%%%%%%%%%%%%%%%%%%%%%%%%%%%%%%%%%%%%%%%%%%%%%%%%%%
%%%%%%%%%%%%%%%%%%%%%%%%%%%%%%%%%%%%%%%%%%%%%%%%%%%%%%%%%%%%%%%
%%%%%%%%%%%%%%%%%%%%%%%%%%%%%%%%%%%%%%%%%%%%%%%%%%%%%%%%%%%%%%%
%  Theory
%%%%%%%%%%%%%%%%%%%%%%%%%%%%%%%%%%%%%%%%%%%%%%%%%%%%%%%%%%%%%%%
%%%%%%%%%%%%%%%%%%%%%%%%%%%%%%%%%%%%%%%%%%%%%%%%%%%%%%%%%%%%%%%
%%%%%%%%%%%%%%%%%%%%%%%%%%%%%%%%%%%%%%%%%%%%%%%%%%%%%%%%%%%%%%%
\section{Theory}
\label{sec:theory}

%%%%%%%%%%%%%%%%%%%%%%%%%%%%%%%%%%%%%%%%%%%%%%%%%%%%%%%%%%%%%%%
%%%%%%%%%%%%%%%%%%%%%%%%%%%%%%%%%%%%%%%%%%%%%%%%%%%%%%%%%%%%%%%
%  Model
%%%%%%%%%%%%%%%%%%%%%%%%%%%%%%%%%%%%%%%%%%%%%%%%%%%%%%%%%%%%%%%
%%%%%%%%%%%%%%%%%%%%%%%%%%%%%%%%%%%%%%%%%%%%%%%%%%%%%%%%%%%%%%%
\subsection{Model}
\label{subsec:model}
The model Hamiltonian for Eu$_{1-x}$Gd$_x$O as well as EuO$_{1-x}$ reads,
\begin{equation}\label{eqn:model_bulk_h}
	H=H_0 + H_{cf} + H_{imp}  \,.
\end{equation}
The conduction band, comprised of the hybridizing Eu $5d\,6s$ 
orbitals, is described by
\begin{equation}\label{eqn:model_bulk_hc}
	H_0=\sum_{\ve{k}\sigma} (\varepsilon_{\ve{k}}-\mu) \cre{c}{\ve{k}\sigma}\ann{c}{\ve{k}\sigma} \,,
\end{equation}
where $\cre{c}{\ve{k}\sigma}$ is the conduction electron creation operator 
and $\varepsilon_{\ve{k}}$ the conduction band dispersion. 
$\mu$ is the chemical potential that fixes the total electron 
number (conduction and impurity electrons). 
In undoped EuO, $\mu$ lies in the gap below the conduction band.
The lattice of Eu $4f$ moments is described by a Heisenberg Hamiltonian and
a local coupling term between the Eu $4f$ moments and the 
conduction electron spins,
\begin{equation}\label{eqn:model_bulk_hcf}
	H_{cf}=- \sum_{\langle i,j \rangle } J_{ij} \ve{S}_{i} \cdot \ve{S}_{j} -
        J_{cf} \sum_i \vs{\sigma}_{i} \cdot \ve{S}_{i} \ .
\end{equation}
Here, $\ve{S}_{i}$ is the $4f$ spin, $m_S=-7/2, \dots , +7/2$, 
and $\vs{\sigma}_i=\frac{1}{2}\sum_{\sigma\sigma'} \cre{c}{i \sigma}
\vs{\tau}_{\sigma\sigma'} \ann{c}{i \sigma'}$ is the conduction
electron spin operator at site $i$.
$J_{ij} >0$ is the direct exchange coupling between the 
localized moments which is independent of the conduction band occupation 
and therefore responsible for the Curie temperature of \SI{69}{K} in 
stoichiometric EuO. $J_{cf}$ is the exchange coupling between the 
$4f$ moment and the conduction electron spin $\vs{\sigma}_{i}$ 
at lattice site $i$. 
The Gd impurities and O vacancies are described as 
Anderson impurities with a single electron binding energy $E_d<0$ and a 
hybridization with the conduction band, $V$, 
\begin{eqnarray}\label{eqn:model_bulk_hcd}
	H_{cd}&=&(E_d-\mu)\sum_{\{j\}\sigma} \cre{d}{j\sigma}
        \ann{d}{j\sigma}\\ 
        &+& V\sum_{\{j\}\sigma}\bigl(  \cre{c}{j\sigma}\ann{d}{j\sigma}+
        \cre{d}{j\sigma}\ann{c}{j\sigma} \bigr)  + U \sum_{\{j\}}
        \cre{d}{j\uparrow} \ann{d}{j\uparrow}\cre{d}{j\downarrow}
        \ann{d}{j\downarrow} \ , \nonumber
\end{eqnarray}
where $\cre{d}{j\sigma}$ is the electron creation operator in an impurity or
defect orbital at site $j$ and $\{j\}$ indicates a summation over the 
randomly placed impurity sites. 
In the following we will use the term impurity for both, Gd impurities and 
O vacancies, unless stated otherwise. The impurity number density will be 
denoted by $n_I$. Gd carries one extra electron in the $5d$ shell as 
compared to Eu. Hence, Gd is in the strongly 
correlated regime with a strong onsite repulsion $U$ preventing 
double occupancy.  
Due to stoichiometry, an O vacancy attracts two extra electrons from the 
surrounding metal ions. Therefore, it is in the weakly correlated 
regime, with double occupancy in the ground state, i.e., a moderate
onsite repulsion $0<U\ll |E_d|$.  
The model, Eqs.~(\ref{eqn:model_bulk_h})--(\ref{eqn:model_bulk_hcd}), inherently
incorporates the RKKY interaction [\onlinecite{RKKY1,RKKY2,RKKY3}] via 2nd-order,
non-local perturbation theory in $J_{cf}$. Since, apart from RKKY effects,
only local self-energies will be important for the following treatment of
this paper, we here include the RKKY Hamiltonian explicitly,
\begin{eqnarray}\label{eqn:HRKKY} 
H^{RKKY} \hspace*{-0.1cm}=\hspace*{-0.08cm}
 - \sum_{(i\neq j)} \left[ K_{ij}^{||}\,  S_i^z S_j^z 
             + K_{ij}^{\perp}\, \left(S_i^xS_j^x + S_i^yS_j^y\right)\right]
 \, .
\end{eqnarray}
It is to be amended to the model Hamiltonian~(\ref{eqn:model_bulk_h}).
A recollection of its derivation and the expressions for the coupling 
constants are given in Appendix \ref{appendixA}, 
see Eqs.~(\ref{eqn:HRKKY3}), (\ref{eqn:KRKKY}).

%%%%%%%%%%%%%%%%%%%%%%%%%%%%%%%%%%%%%%%%%%%%%%%%%%%%%%%%%%%%%%%
%%%%%%%%%%%%%%%%%%%%%%%%%%%%%%%%%%%%%%%%%%%%%%%%%%%%%%%%%%%%%%%
%  Selfconsistent theory
%%%%%%%%%%%%%%%%%%%%%%%%%%%%%%%%%%%%%%%%%%%%%%%%%%%%%%%%%%%%%%%
%%%%%%%%%%%%%%%%%%%%%%%%%%%%%%%%%%%%%%%%%%%%%%%%%%%%%%%%%%%%%%%
\subsection{Selfconsistent theory}
\label{subsec:selfconsistent_theory}

To evaluate this model, we follow and extend Ref.~[\onlinecite{Arnold08}].
While the large spins of the $4f$ Heisenberg lattice can be treated in 
mean-field theory, it is essential to describe the Anderson impurities
dynamically, in order to account for their spin and charge fluctuations.
The conduction electron selfenergy induced by the impurities will be 
treated in a single-site approximation, i.e. it is given by the 
full impurity $T$ matrix times the impurity concentration $n_I$. This is valid 
for dilute impurities, where inter-impurity correlations are negligible. 
Writing the (retarded, $\omega\equiv\omega+i0$) 
conduction electron Green's function as, 
\begin{equation}\label{eqn:tcb_bulk_green}
G_{c\sigma}(\ve{k}, \omega)=
\frac{1}{\omega+\mu-\varepsilon_{\ve{k}}-\Sigma_{c\sigma}(\omega)} 
\end{equation}
yields for the total conduction selfenergy,
\begin{equation}\label{eqn:Sigma_c}
\Sigma_{c\sigma} (\omega) = n_I V^2 G_{d\sigma}(\omega) -\sigma J_{cf}\ex{S}\,,
\quad \sigma=\pm \frac{1}{2} \, ,
\end{equation}
where $G_{d\sigma}(\omega)$ is the full Green's function of the impurity 
electrons. 
The expectation value of the $4f$ spins is determined in mean-field theory
by (with $\beta=1/k_BT$, the inverse temperature),
\begin{equation}
	\ex{S}=\frac{\sum_{S=-\frac{7}{2}}^\frac{7}{2} S e^{\beta(2 J_{4f}
            \ex{S} + J_{cf}\ex{\sigma})S}}{\sum_{S=-\frac{7}{2}}^\frac{7}{2}
          e^{\beta(2 J_{4f} \ex{S} +
            J_{cf}\ex{\sigma})S}} \label{eqn:tcf_meanfield} \,,
\end{equation}
\begin{figure}[t]
\begin{centering}
 \includegraphics[width=1.0\linewidth]{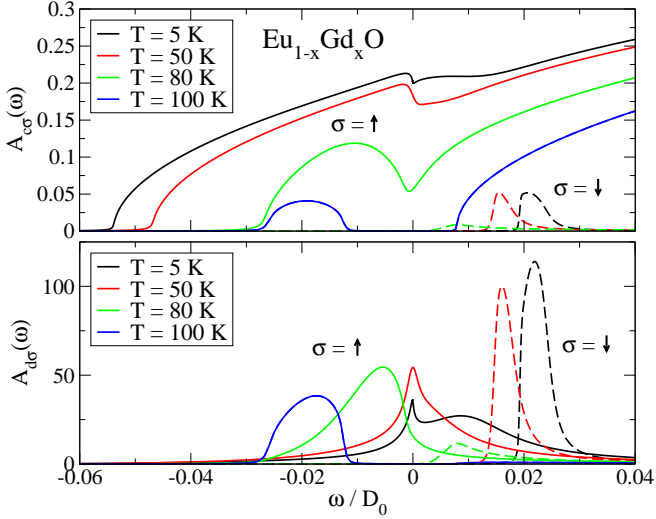}
\end{centering}
\vspace*{-0.32cm}
 \caption{(Color online) Conduction band (upper panel) and Gd impurity 
(lower panel) spectral density across the phase transition in 
Eu$_{1-x}$Gd$_{x}$O for $x=0.01$. $T_C\approx 95$~K. 
Solid curves represent the majority 
($\sigma = \uparrow$) and dashed curves the minority ($\sigma = \downarrow$)
spin spectral density.}
\label{fig:Gd_Ad_Ac_across_phase_transition}
\end{figure}

Here, the conduction electron magnetization $\ex{\sigma}$ is calculated as,
\begin{equation}
\ex{\sigma}=\frac{1}{2}\int d\omega
f(\omega)\left[A_{c\uparrow}(\omega)-A_{c\downarrow}(\omega)\right] \,,
\label{eqn:avg_spin}
\end{equation}
with $A_{c\sigma}(\omega)=-\frac{1}{\pi}\sum_{\ve{k}}\im
G_{c\sigma}(\ve{k},\omega+i0)$ the spin-dependent, interacting conduction 
electron density of states (DOS). $J_{4f}=\sum_j J_{0j}$ is the effective
mean-field coupling of the Heisenberg lattice. The short-range nature 
of $J_{ij}$ restricts the lattice summation essentially to the nearest neighbors of 
site $i=0$. The magnetic polaron effect [\onlinecite{Mauger77,Mauger86}] is
incorporated in Eq.~(\ref{eqn:tcf_meanfield}) via the 
conduction electron magnetization $\ex{\sigma}$.
For later use, the conduction carrier density is given by,
$n_c= \sum_\sigma \int d\omega f(\omega) A_{c\sigma}(\omega)$. 
The calculation of the local impurity Green's function 
$G_{d\sigma}(\omega)$ depends on whether the impurity is in the strongly 
(Gd impurities) or the weakly (O vacancies) correlated regime.\\

\begin{figure}[t]
\begin{centering}
 \includegraphics[width=0.98\linewidth]{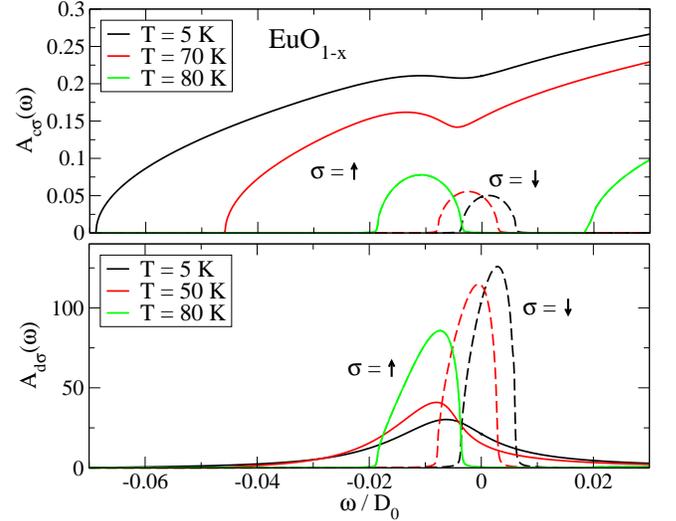}
\end{centering}
 \caption{(Color online)
Conduction band (upper panel) and O vacancy 
(lower panel) spectral density across the phase transition
 in EuO$_{1-x}$ for $x=0.01$. The bare O-defect parameters are 
$E_d=0.0\,D_0$, $U=0$ (double occupancy). $T_C\approx 78$~K. 
Solid curves represent the majority 
($\sigma = \uparrow$) and dashed curves the minority ($\sigma = \downarrow$)
spin spectral density.}
\label{fig:O_Ad_Ac_across_phase_transition}
\end{figure}

$a$. {\it Gd impurities}. 
We employ the auxiliary particle technique in non-crossing approximation 
[\onlinecite{Keiter81,Kuramoto83,Coleman84}] 
to describe the spin and charge fluctuations in the  Gd $5d$ orbitals.
The limit $U\to\infty$ is taken, for simplicity, in order to prevent 
double occupancy.  
Since in Eu$_{1-x}$Gd$_x$O the DOS near the chemical 
potential is low or even vanishing, the Kondo temperature of the 
fluctuating spin, $T_K$, is far below $T_C$. The NCA is known to give 
reliable results for energies above and down to somewhat below $T_K$.
In a magnetic field it produces, in addition to the 
Zeeman-split Kondo resonance, a spurious low-temperature 
singularity at the Fermi level for $T<T_K$. However, since $T_K\ll T_C$ 
in Eu$_{1-x}$Gd$_x$O, the effect of this singularity is negligible for the 
temperature range relevant here.
The NCA is also versatile enough to include an arbitrary energy dependence 
of the DOS. Therefore, it is the appropriate method for the 
present purpose [\onlinecite{Arnold08}]. 
The NCA equations, adapted for the Eu$_{1-x}$Gd$_x$O system with a 
gapped DOS and a non-trivial chemical potential, 
are given in Appendix \ref{appendixB}.

$b$. {\it O vacancies}. 
The weak interaction on the O defects, where spin fluctuations are 
negligible, can be accounted for in the second-order perturbation theory in $U$. 
The (retarded) O-defect electron Green's function is 
$G_{d\sigma}(\omega)=1/\left[\omega+\mu-E_d-\Sigma_{d\sigma}(\omega)\right]$,
and the corresponding selfenergy reads, 
$\Sigma_{d\sigma}(\omega)=\Sigma_{d\sigma}^{(1)}(\omega) + 
\Sigma_{d\sigma}^{(2)}(\omega)$, where
\begin{eqnarray}
\Sigma_{d\sigma}^{(1)}(\omega) &=& \pi V^2 A_{c\sigma}(\omega) 
    + U \int d \varepsilon f(\varepsilon) A_{d-\sigma}(\varepsilon) \\
{\rm Im}\Sigma_{d\sigma}^{(2)}(\omega) &=& \\ 
&&\hspace*{-2cm} - U^2  \int d \varepsilon_1 \int d \varepsilon_2\,   
A_{d\sigma}(\varepsilon_1+\omega)A_{d-\sigma}(\varepsilon_1+\varepsilon_2)A_{d-\sigma}(\varepsilon_2) \nonumber\\
&& \hspace*{0.1cm} \times  
  \left[ b(\varepsilon_1)+f(\varepsilon_1+\omega) \right] 
  \left[ f(\varepsilon_2)-f(\varepsilon_1+\varepsilon_2)\right]\,,
\nonumber
\end{eqnarray}
where $f(\omega)$ and $b(\omega)$ are Fermi and Bose distribution 
functions, respectively, and ${\rm Re}\Sigma_{d\sigma}^{(2)}(\omega)$ 
is given by the Kramers-Kronig relation.
  
\begin{figure}[b] 
\begin{centering}
 \includegraphics[angle=0,width=\linewidth]{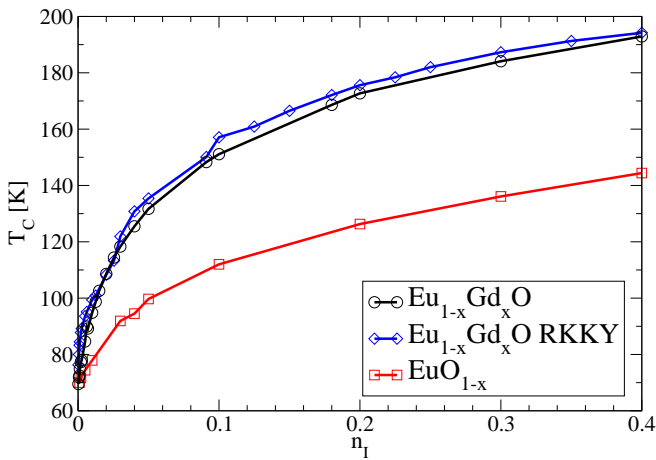}
\end{centering}
\vspace*{-0.4cm}
\caption{(Color online) Curie temperature vs.\ the doping concentration $n_I$ 
for Gd-doped 
and O-deficient EuO. Bare parameters for O vacancies: $E_d=0.0\,D_0$, $U=0$.
The blue curve represents the behavior for Eu$_{1-x}$Gd$_x$O including the
RKKY interaction, see text.}
\label{fig:tc_vs_ni}
\end{figure}

{\it Long-range RKKY coupling}. When the RKKY interaction is included, the 
$4f$ Heisenberg coupling is changed to $J_{ij}\to J_{ij} + K_{ij}$. 
In the above equations this leads to a modified mean-field coupling,
\begin{eqnarray}
    J_{4f}  = \sum_{j n.n. 0} J_{0j} + \sum_{j\neq 0} K_{0j}^{||} \, .
 \label{eqn:tcf_rkky_bulk_coupling}
\end{eqnarray}
The lattice summation in the second term is carried out 
over the fcc lattice of the EuO rocksalt structure.
Note that on mean-field level only the longitudinal RKKY component 
contributes and can give FM as well as AF contributions to the 
total coupling. $K_{0j}^{||}$ involves non-local 
Green's functions (c.f. Appendix \ref{appendixA}) and, hence, 
the band dispersion $\varepsilon_{\bf k}$. For simplicity and since anisotropy 
effects are not important in bulk EuO, we assume for the RKKY interaction 
an isotropic dispersion which is constructed such [\onlinecite{Kroha90}] that it 
reproduces the bare conduction DOS.  
 
The system is subject to the doping condition that the total density 
of charge carriers is $n_{tot}=n_I$ for Eu$_{1-x}$Gd$_x$O and 
$n_{tot}=2\,n_I$ for EuO$_x$. That is, 
\begin{equation} \label{eqn:charge_carrier_constraint}
    \sum_\sigma \int d\omega f(\omega) \left[ A_{c\sigma}(\omega) + n_I
      A_{d\sigma}(\omega)\right] - n_{tot} =0  \, .
\end{equation}
The selfconsistent set of equations 
(\ref{eqn:tcb_bulk_green})--(\ref{eqn:charge_carrier_constraint}), in the
case of Eu$_{1-x}$Gd$_x$O amended by the NCA equations 
(\ref{eqn:nca_sigma_f})--(\ref{eqn:nca_green_d}), 
is solved by iteration, where in each iteration step the chemical 
potential $\mu$ is adjusted so as to fulfill the 
particle number constraint (\ref{eqn:charge_carrier_constraint}). 
Note that the RKKY coupling strength $K_{ij}^{||}$, Eq.~(\ref{eqn:KRKKY}),
is also determined selfconsistently via the interacting  
conduction electron propagators.

%%%%%%%%%%%%%%%%%%%%%%%%%%%%%%%%%%%%%%%%%%%%%%%%%%%%%%%%%%%%%%%
%%%%%%%%%%%%%%%%%%%%%%%%%%%%%%%%%%%%%%%%%%%%%%%%%%%%%%%%%%%%%%%
%%%%%%%%%%%%%%%%%%%%%%%%%%%%%%%%%%%%%%%%%%%%%%%%%%%%%%%%%%%%%%%
%  Results
%%%%%%%%%%%%%%%%%%%%%%%%%%%%%%%%%%%%%%%%%%%%%%%%%%%%%%%%%%%%%%%
%%%%%%%%%%%%%%%%%%%%%%%%%%%%%%%%%%%%%%%%%%%%%%%%%%%%%%%%%%%%%%%
%%%%%%%%%%%%%%%%%%%%%%%%%%%%%%%%%%%%%%%%%%%%%%%%%%%%%%%%%%%%%%%
\section{Results}
\label{sec:results}

\begin{figure}
\begin{centering} 
    \includegraphics[width=\linewidth]{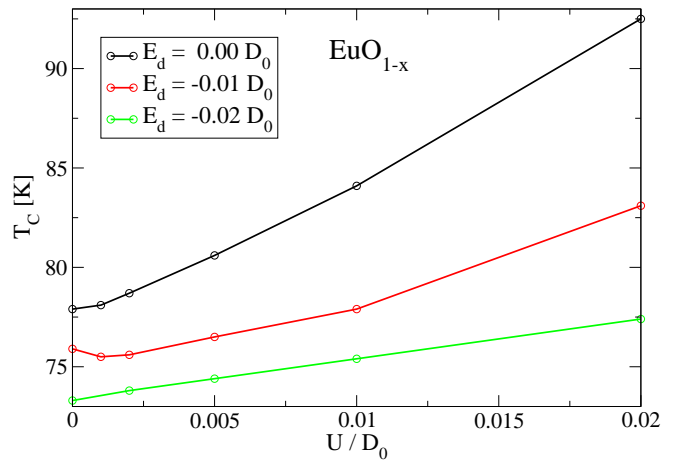}
\end{centering}
\vspace*{-0.3cm}
    \caption{(Color online) Curie temperature vs.\ on-site Coulomb repulsion $U$ in
      EuO$_{0.99}$ for various impurity level energies $E_d$. 
      It is seen that for more tightly bound defect electrons the 
      $T_C$ enhancement in EuO$_{a-x}$ is even weaker than for the parameter 
      values of Fig.~\ref{fig:tc_vs_ni}.}
\label{fig:oxy_tc_vs_U_and_Ed}
\end{figure}

\subsection{Parameter values}.
For the numerical calculations below, we choose a semi-elliptic 
DOS for the non-interacting conduction band of stoichiometric 
EuO, $N_{c\sigma}^{(0)}(\omega)= (2/\pi D_0)\,\sqrt{(\omega-\mu_0-D_0)^2/D_0^2-1}$. 
The conduction half-bandwidth is taken to be $D_0=8$~eV, and 
the chemical potential of the undoped system lies 
in the gap below the conduction band, $\mu_0 = -0.02\, D_0$,  
consistent with experiments [\onlinecite{Steeneken02}]. All energies are
measured relative to the (interacting) chemical potential $\mu$ and 
are given in units of $D_0$. The mean-field Heisenberg coupling $J_{4f}$ 
(without RKKY interaction) is chosen such that $T_C=69$~K is obtained 
for undoped EuO. This yields $J_{4f}=7\cdot 10^{-5}\,D_0$ [\onlinecite{Arnold08}].
$J_{cf}$ is much larger than $J_{4f}$, because the overlap between the 
neighboring Eu $4f$ orbitals is much smaller than their overlap 
with the conduction wave functions. From the spatial separation of the 
Eu $4f$ orbitals the ratio $J_{cf}/J_{4f}$ is 
roughly estimated to give $J_{cf}=0.05\,D0$, see also Ref.~[\onlinecite{Arnold08}].
This also determines the RKKY coupling strength via Eq.~(\ref{eqn:KRKKY}).  
We fix the bare parameters of a Gd impurity such that for $T=0$ and
vanishing impurity concentration its occupation is $n_d\approx 1$ 
and that the impurity electron gets thermally activated in the experimentally 
relevant temperature range. This yields, 
$E_d = 0.0$, $\Gamma := \pi V^2/D_0= 0.05\,D_0$, and $U\to\infty$. 
Note that hybridization and interaction substantially 
renormalize the impurity level, $E_d \to\tilde E_d\approx -0.02\,D_0$ 
(Haldane shift [\onlinecite{Haldane78}], see also 
Fig.~\ref{fig:Gd_Ad_Ac_across_phase_transition}), 
so that $n_d\approx 1$ is realized in the $n_I\to 0$ limit. 
This also renders the $T_K$ of the impurity far below 
$T_C$, since in our system the DOS at the Fermi level $E_F$ remains always 
$A_{c\sigma}(0)\ll 1/D_0$. For O vacancies, in absence of more detailed
information about their structure other than double occupancy, we 
set the effective hybridization $\Gamma = 0.05\,D_0$, as for Gd, 
and perform a scan of $E_d\leq0$ and $U$ within the bound-state, 
weakly correlated regime, see below. The RKKY interaction will be
included and discussed in Subsec.~\ref{subsec:rkky} only.

\begin{figure}[t]
\begin{centering}
\includegraphics[width=\linewidth]{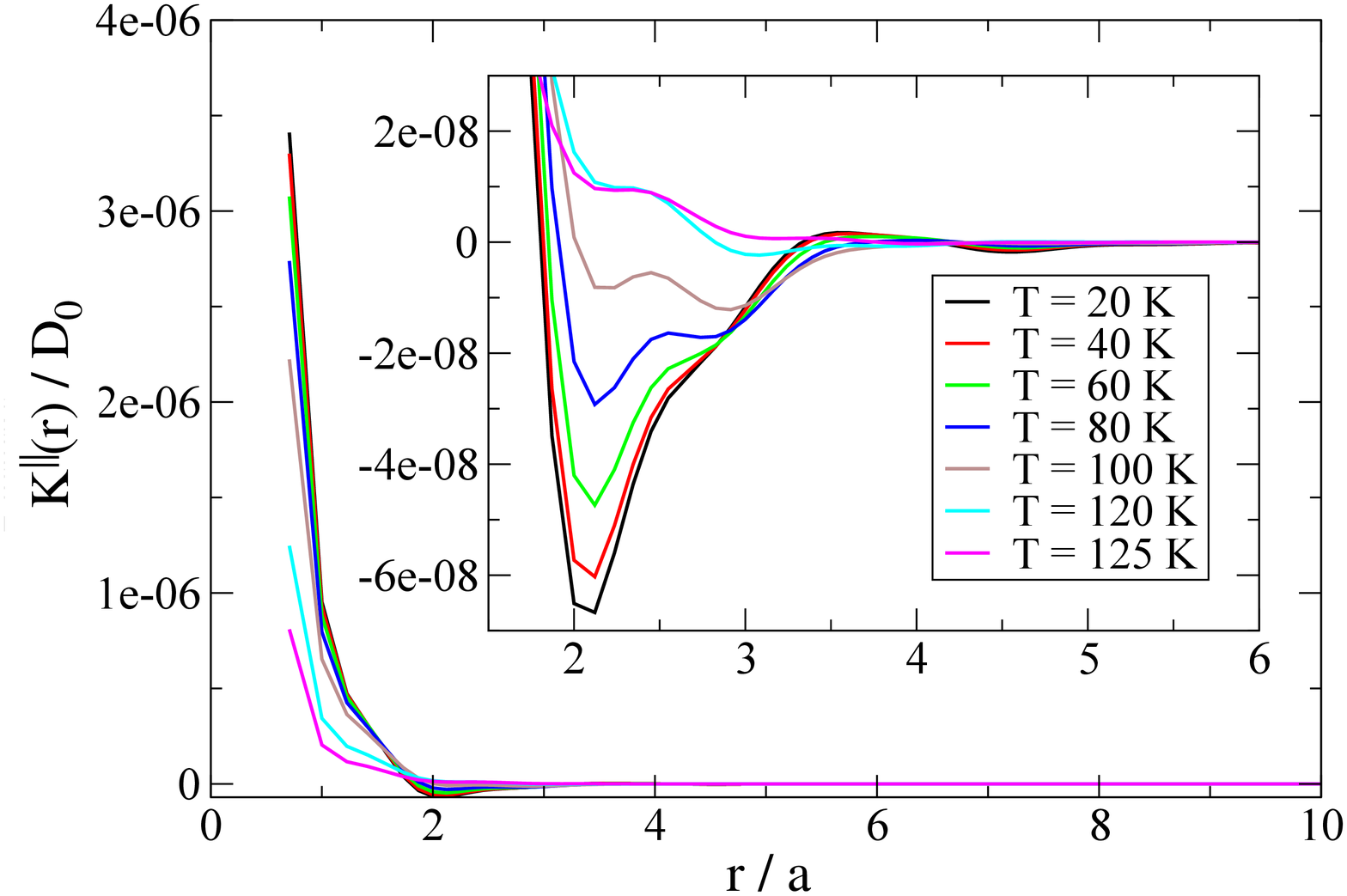}
\end{centering}
\vspace*{-0.7cm}
\caption{(Color online) Longitudinal RKKY coupling $K^{||}(r)$ for 
$x=n_I=0.04$ and various temperatures in Eu$_{1-x}$Gd$_x$O. The inset 
expands the AF region around two to four lattice spacings a of the FCC 
lattice.}
\label{fig:bulk-eugdo-lr-coupling}
\end{figure}

%%%%%%%%%%%%%%%%%%%%%%%%%%%%%%%%%%%%%%%%%%%%%%%%%%%%%%%%%%%%%%%
%%%%%%%%%%%%%%%%%%%%%%%%%%%%%%%%%%%%%%%%%%%%%%%%%%%%%%%%%%%%%%%
%  Gadolinium impurites vs Oxygen vacancies
%%%%%%%%%%%%%%%%%%%%%%%%%%%%%%%%%%%%%%%%%%%%%%%%%%%%%%%%%%%%%%%
%%%%%%%%%%%%%%%%%%%%%%%%%%%%%%%%%%%%%%%%%%%%%%%%%%%%%%%%%%%%%%%
\subsection{Gadolinium impurities vs. Oxygen vacancies}
\label{subsec:Gd_vs_O}

Figures\ref{fig:Gd_Ad_Ac_across_phase_transition} and 
\ref{fig:O_Ad_Ac_across_phase_transition} show the evolution of the 
conduction band and impurity spectral densities across the 
phase transition for low-doped Eu$_{1-x}$Gd$_{x}$O and EuO$_{1-x}$,
respectively. For both, Eu$_{1-x}$Gd$_{x}$O and EuO$_{1-x}$,
in the high-temperature insulating phase the spin degenerate conduction DOS 
is comprised of a large, unoccupied band and a small side band which is 
induced by the hybridization with the impurity orbitals and is 
centered around the impurity binding energy $E_d$, lying entirely 
below $\mu$ and, therefore, not contributing to the conductivity.
As the temperature is lowered, in Eu$_{1-x}$Gd$_{x}$O the impurity spectrum
accumulates spectral weight at the chemical potential which eventually 
develops into a peak at $\omega=0$ 
(Fig.~\ref{fig:Gd_Ad_Ac_across_phase_transition}). Below $T_C$ the 
spectral densities are split into majority and minority bands.
Our NCA calculations show that this is due to local, Kondo-like spin 
fluctuations in the Gd $5d$ orbitals [\onlinecite{Arnold08}]. Because of hybridization, 
the conduction electron DOS develops spectral weight at $\omega=0$ as well, 
and the side band merges with the main conduction band. 
This drives the metallic transition and simultaneously enhances, 
via the magnetic polaron effect [c.f. Eq.~(\ref{eqn:tcf_meanfield})], 
the FM transition temperature as well. In EuO$_{1-x}$, the local spin 
fluctuation effect is absent. Here, the metallic transition occurs only 
when the conduction side band is eventually broadened and shifted, via 
hybridization with the O vacancy band enough to gain overlap with 
the chemical potential (Fig.~\ref{fig:O_Ad_Ac_across_phase_transition}), 
leading to a much lower $T_C$ than in Eu$_{1-x}$Gd$_{x}$O.
In Fig.~\ref{fig:tc_vs_ni} the doping-dependent $T_C$ enhancement is 
compared for magnetic Gd impurities and non-magnetic O vacancies
(black and red curves). 
Here, for O vacancies, $U=0$ (double occupancy) and otherwise the same parameter
values as for Eu$_{1-x}$Gd$_x$O are assumed. This allows for a direct 
assessment of the importance of low-lying, local spin fluctuations 
for the $T_C$ enhancement.
The essential role of on-site correlations as well as conduction electron
doping is further substantiated by Fig.~\ref{fig:oxy_tc_vs_U_and_Ed}, where
$T_C$ is shown for varying $U$ and $E_d$ values in EuO$_{1-x}$: 
$T_C$ is enhanced by repulsive onsite correlations (increasing $U$) and 
is reduced by the dopant electrons more tightly bound to the defect 
(more negative $E_d$).

%%%%%%%%%%%%%%%%%%%%%%%%%%%%%%%%%%%%%%%%%%%%%%%%%%%%%%%%%%%%%%%
%%%%%%%%%%%%%%%%%%%%%%%%%%%%%%%%%%%%%%%%%%%%%%%%%%%%%%%%%%%%%%%
%  RKKY coupling
%%%%%%%%%%%%%%%%%%%%%%%%%%%%%%%%%%%%%%%%%%%%%%%%%%%%%%%%%%%%%%%
%%%%%%%%%%%%%%%%%%%%%%%%%%%%%%%%%%%%%%%%%%%%%%%%%%%%%%%%%%%%%%%
\subsection{RKKY interaction in Eu$_{1-x}$Gd$_x$O}
\label{subsec:rkky}

\begin{figure}
\begin{centering}
\includegraphics[width=0.92\linewidth]{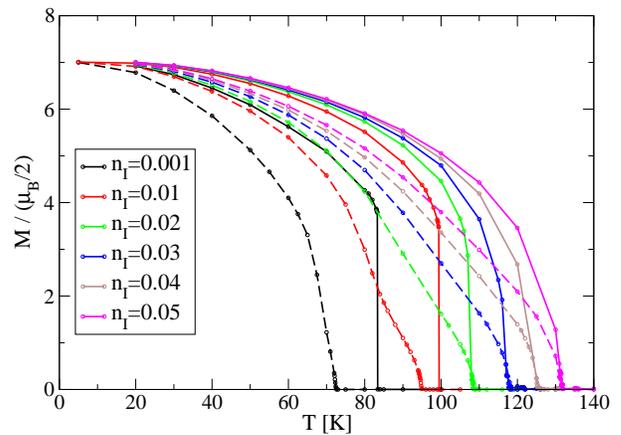}
\end{centering}
    \caption{(Color online) Total magnetization $M=\ex{S}+\ex{\sigma}$ vs.\ temperature $T$ for various doping concentrations in Eu$_{1-x}$Gd$_x$O with long range RKKY coupling (solid curves) and without RKKY coupling (dashed curves).}
	\label{fig:bulk-eugdo-lr-mag_vs_ni}
\end{figure}

We now study the influence of the long-range 
RKKY interaction on the phase transition in Eu$_{1-x}$Gd$_x$O. 
For short Fermi wavelength, the RKKY interaction might make an 
AF contribution to the total coupling and, thus, lead to the 
experimentally observed saturation of $T_C$
[\onlinecite{Schoenes74,Matsumoto04,Ott06,Mairoser10,Altendorf12,Mairoser13}], 
as has been suggested in Ref.~[\onlinecite{Kogan11}].  
In order to analyze the possible influence of the RKKY interaction on the 
saturation at high doping concentration, we now adjust the value of $J_{cf}$ 
such that the theory {\it including} RKKY reproduces the previous 
results {\it without} RKKY interaction (Sec. \ref{subsec:Gd_vs_O}) 
in the low-doping regime, and will compare the results at high doping. 
This yields the new value $J_{cf}=0.0405\, D_0$. 
The RKKY coupling $K^{||}(r)$, selfconsistently calculated for the interacting
system, is displayed in Fig.~\ref{fig:bulk-eugdo-lr-coupling}  
as a function of distance $r$ for a typical Gd doping concentration 
of $n_I=0.04$ over the complete, relevant $T$ range. It shows weak AF  
behavior only in the range of about 2 to 4 FCC lattice constants. 
The resulting total magnetization $M$ is shown in 
Fig.~\ref{fig:bulk-eugdo-lr-mag_vs_ni}. While the FM magnetization is 
even enhanced by $K^{||}$ below the transition, 
it does not substantially alter $T_C$, especially for higher doping. 
This is plausible, because
the RKKY interaction is not active for $T>T_C$ (empty conduction band),
but its long-range, overall FM behavior enhances $M$ once the 
band is populated for $T<T_C$. Such enhancement of the FM coupling by 
an RKKY-like interaction is consistent with recent experiments 
on EuO doped with non-magnetic La atoms [\onlinecite{Monteiro15}].
Fig.~\ref{fig:tc_vs_ni} directly
compares $T_C$ with and without RKKY coupling (blue and black curves, 
respectively) in our calculation. 
It is seen that including the RKKY interaction and 
reducing the direct exchange coupling to $J_{cf}=0.0405\, D_0$
not only reproduces the $T_C$ behavior at small $n_I$ 
(by construction), but also does not change the behavior for the
largest $n_I$ considered. For the small band fillings relevant in
Eu$_{1-x}$Gd$_x$O the effects of the RKKY interaction are 
essentially doping independent and can be absorbed in a proper 
choice of $J_{cf}$, at least as far as $T_C$ in bulk systems is concerned.
We conclude that the experimentally observed $T_C$ saturation behavior 
in Eu$_{1-x}$Gd$_x$O for large $n_I$ cannot be attributed to the RKKY 
interaction.
Note that {\it ab initio} calculations 
[\onlinecite{Belashchenko13,Schwingenschloegl15}] presumably overestimate the 
antiferromagnetic contributions from the RKKY interaction, because they do 
not take the Kondo-like spin fluctuations on the Gd sites and the resulting 
accumulation of spectral weight at the chemical potential into account. 
As a consequence, the RKKY wavelength comes out too short and, hence, 
its antiferromagnetic contributions too large. 
This may be the origin why these calculations overestimate the decrease 
of $T_C$ for large doping concentration as compared to 
experiments [\onlinecite{Mairoser10,Altendorf12}].

\begin{figure}[t]
\begin{centering}
 \includegraphics[width=\linewidth]{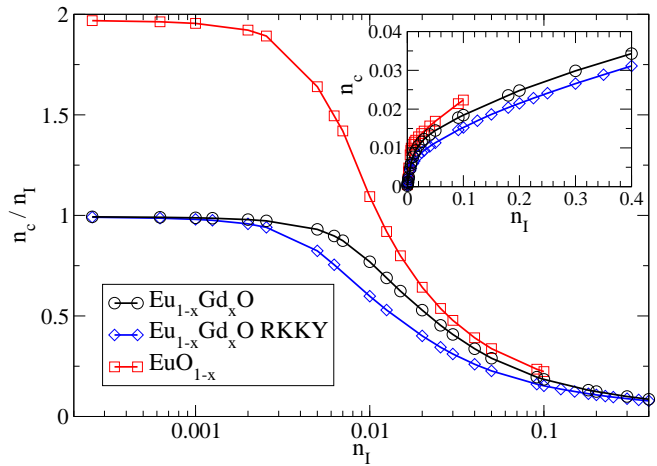}
\end{centering}
\vspace*{-0.3cm}
 \caption{(Color online) Dopant activation $n_c/n_I$ for Gd doped and O deficient EuO at
the lowest temperature considered, $T=5 K$.
The carrier concentration $n_c$ is shown in the inset.}
\label{fig:nc_vs_ni}
\end{figure}

%%%%%%%%%%%%%%%%%%%%%%%%%%%%%%%%%%%%%%%%%%%%%%%%%%%%%%%%%%%%%%%
%%%%%%%%%%%%%%%%%%%%%%%%%%%%%%%%%%%%%%%%%%%%%%%%%%%%%%%%%%%%%%%
%  Dopant activation
%%%%%%%%%%%%%%%%%%%%%%%%%%%%%%%%%%%%%%%%%%%%%%%%%%%%%%%%%%%%%%%
%%%%%%%%%%%%%%%%%%%%%%%%%%%%%%%%%%%%%%%%%%%%%%%%%%%%%%%%%%%%%%%
\subsection{Dopant activation and $T_C$ saturation}
\label{subsec:DopantActivation} 

\begin{figure}
\begin{centering}
 \includegraphics[angle=0,width=0.87\linewidth]{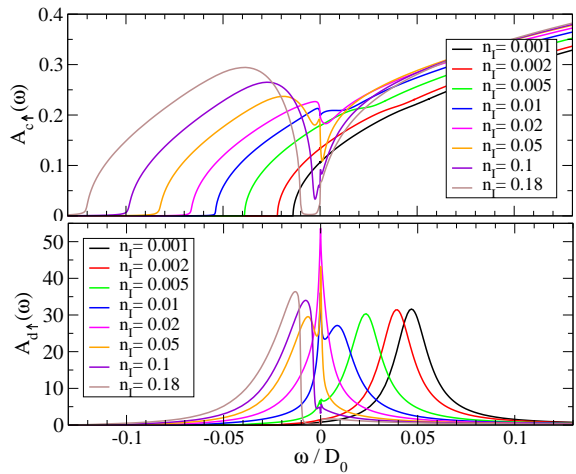}
\end{centering}
\vspace*{-0.1cm}
\caption{(Color online) Majority spectral densities for conduction electrons (upper panel) 
and impurity electrons (lower panel) in Eu$_{1-x}$Gd$_x$O at $T=5$~K for 
various impurity concentrations $n_I$. The figure shows the downward 
shift of the impurity levels with increasing $n_I$.}
\label{fig:aduacu_vs_ni}
\end{figure}

\begin{figure}[t]
\begin{centering}
 \includegraphics[angle=0,width=\linewidth]{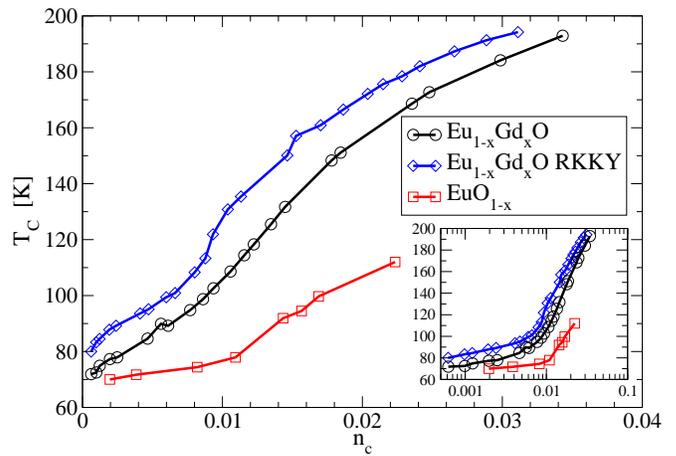}
\end{centering}
\vspace*{-0.3cm}
\caption{(Color online) Curie temperature vs.\ carrier concentration $n_c$ in the 
conduction band for Gd-doped and O-deficient EuO. 
(Inset) Semilogarithmic plot of the data.}
\label{fig:tc_vs_nc}
\end{figure}

In Fig.~\ref{fig:nc_vs_ni} we show the charge carrier concentration
in the conduction band $n_c$ (number of carriers per lattice unit cell; inset) 
as well as the dopant activation $n_c/n_I$  as a function of 
impurity concentration $n_I$ at the lowest temperature considered,
$T=5$~K, similar to Ref.~[\onlinecite{Mairoser10}]. 
For low doping the impurity spectral weight 
$A_{d\sigma}(\omega)$ (both, $\sigma =\uparrow,\, \downarrow$) 
is almost entirely shifted above $E_F$ at this temperature,
as can bee seen from Figs.~\ref{fig:Gd_Ad_Ac_across_phase_transition}, 
\ref{fig:O_Ad_Ac_across_phase_transition}, lower panels, so that the
impurity level is completely emptied into the conduction band. 
Consequently, the dopant activation is $n_c/n_I=1$  (Gd) or
$n_c/n_I=2$ (O vacancies) up to a doping concentration of about 
$n_I=0.01$ (Fig.~\ref{fig:nc_vs_ni}). For higher $n_I$, the impurity 
contribution to the conduction electron selfenergy, Eq.~(\ref{eqn:Sigma_c}),
gets increasingly enhanced by the disorder. Via the Kramers-Kronig 
relation for the real part of $\Sigma_{c\sigma}(\omega)$ this implies a 
downward shift of the conduction side band and, connected with it, 
a downward shift of the impurity band below $E_F$. This is 
seen in Fig.~\ref{fig:aduacu_vs_ni}. It results in a re-population of 
the impurity levels and a crossover to a reduced $n_c/n_I$, 
as seen in Fig.~\ref{fig:nc_vs_ni}. Note that the description of the 
Gd impurities as Anderson impurities with spin fluctuations is crucial 
for the downward shift of the impurity levels. 
The reduction of the dopant activation $n_c/n_I$ is in agreement with 
the experimental findings of Ref.~[\onlinecite{Mairoser10}]. 
Note that in Ref.~[\onlinecite{Mairoser10}]
a reduced dopant activation is also found in the limit of {\it small} 
$n_I$. Presumably this is, because their Hall measurements determine
the mobile carrier density $n$ only. However, for small impurity 
concentration, part of the electrons in the conduction band will 
be bound around the impurity locations. However, all electrons in the 
conduction band, given by $n_c$, contribute to the electron-enhanced 
magnetic coupling. Comparing the doping dependence of $n_c$ in the
inset of Fig.~\ref{fig:nc_vs_ni} with $T_C$ in Fig.~\ref{fig:tc_vs_ni} 
shows that the latter follows the behavior of $n_c$. 
Displaying now $T_C$ (same data as in Fig.~\ref{fig:tc_vs_ni}) 
as a function of the carrier concentration $n_c$ in 
Fig.~\ref{fig:tc_vs_nc} shows that it grows essentially linearly with 
$n_c$, showing only a slightly decreasing slope for the highest 
$n_c$. Note that the highest carrier concentration of $n_c\approx 0.04$ appears
experimentally achievable, while the corresponding doping concentration 
of $n_I=0.4$ is not, due to stability reasons of the crystal structure. 
The saturation-like behavior of $T_C$ for large $n_I$ doping is, 
therefore, to be considered a consequence of the reduced dopant activation 
for large doping, in complete agreement with the conclusion 
of Ref.~[\onlinecite{Mairoser10}].

%%%%%%%%%%%%%%%%%%%%%%%%%%%%%%%%%%%%%%%%%%%%%%%%%%%%%%%%%%%%%%%
%%%%%%%%%%%%%%%%%%%%%%%%%%%%%%%%%%%%%%%%%%%%%%%%%%%%%%%%%%%%%%%
%%%%%%%%%%%%%%%%%%%%%%%%%%%%%%%%%%%%%%%%%%%%%%%%%%%%%%%%%%%%%%%
%  Conclusion
%%%%%%%%%%%%%%%%%%%%%%%%%%%%%%%%%%%%%%%%%%%%%%%%%%%%%%%%%%%%%%%
%%%%%%%%%%%%%%%%%%%%%%%%%%%%%%%%%%%%%%%%%%%%%%%%%%%%%%%%%%%%%%%
%%%%%%%%%%%%%%%%%%%%%%%%%%%%%%%%%%%%%%%%%%%%%%%%%%%%%%%%%%%%%%%

\section{Conclusion}
\label{sec:conclusion}
We have performed a detailed comparison of the FM insulator-metal transitions
in Eu$_{1-x}$Gd$_x$O and in EuO$_{1-x}$, respectively, using a model that
treats the dopant impurities as Anderson impurities in the strongly (Gd) or 
weakly (O vacancies) correlated regime, and that had previously provided a 
detailed description [\onlinecite{Arnold08}] of experimental magnetization, 
resistivity and total conduction band polarization data. 
Our results show that for a significant, doping-induced 
$T_C$ enhancement a cooperation of two effects is necessary, (1) 
Kondo-like, low-energy spin fluctuations accumulating impurity 
as well as conduction spectral weight at the Fermi energy and 
(2) efficient population of this low-lying spectral weight and subsequent 
enhancement of the FM interaction between the $4f$ moments mediated by the 
conduction electrons. In addition, our calculations provide evidence that
the tendency of $T_C$ to saturate for high doping concentrations is not 
due to AF contributions of the RKKY interaction but rather due to a limitation
of the dopant electron activation into the conduction band, confirming 
experimental results [\onlinecite{Mairoser10}]. Hence, an increase of $T_C$ 
beyond the presently achievable values may be possible, if only the 
conduction band can be populated in a more efficient way. This is in line 
with recent pump-probe experiments [\onlinecite{Matsubara15}] where 
enhanced FM coupling was achieved by photodoping into the conduction
band [\onlinecite{footnote1}].
The combination of all these findings point to a possible pathway to further 
enhancement of $T_C$: the magnetic impurities generating low-energy 
spin fluctuations and the charge-doping impurities need not necessarily
be of the same type. More efficient carrier doping may be achievable 
by using different types of donor atoms (with impurity levels closer or 
above the Fermi energy) in addition to Gd, or by 
carrier coupling at interfaces in heterostructures.

\section{Acknowledgments}
We gratefully acknowledge useful discussions with Carsten Busse, 
Manfred Fiebig, J\"urgen Klinkhammer, Jochen Mannhart, Andreas Schmehl 
and Hao Tjeng. This work was supported in part by the Deutsche 
Forschungsgemeinschaft through SFB 608.

\appendix

\section{RKKY interaction in paramagnetic metals and semimetals} 
\label{appendixA}

\begin{figure}
\begin{centering}
\includegraphics[width=0.5\linewidth]{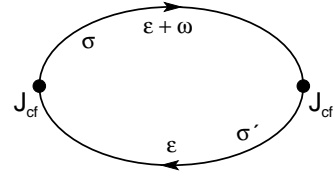}
\end{centering}
\caption{
Diagram for the spin-dependent conduction electron polarization function 
$\Pi_{ij}^{\sigma\sigma'}(\omega)$, generating the RKKY interaction.
The solid lines represent conduction electron propagators.
\label{fig:diagram_RKKY}
}
\vspace*{-0.3cm}
\end{figure}

In this appendix we derive the expressions for the RKKY interaction,
allowing for an arbitrary magnetization of the conduction band 
and then specializing for the paramagnetic case (vanishing magnetization)
and the semimetallic case (complete magnetization). The interaction Hamiltonian 
between localized spins $\ve{S}_i$ at sites $i$ and the conduction electron 
spins reads,    
\begin{eqnarray}\label{eqn:Hcf}
H_{cf}= - J_{cf} \sum_i\ve{S}_{i}\cdot \ve{s}_{i} \ ,
\end{eqnarray}
where $\ve{s}_{i}=1/2 \ \sum_{\sigma\sigma'} 
c_{i\sigma}^{\dagger}\vs{\sigma}c_{i\sigma '}^{\phantom{\dagger}}$ is the conduction 
electron spin operator at site $i$ and 
$\vs{\sigma}=(\sigma^x,\sigma^y,\sigma^z)$ the vector of Pauli matrices.
Evaluating the time evolution of the conduction electrons in the 
presence of another localized spin $\ve{S}_j$ according to 
${\rm exp}[-\int_0^\beta d\tau H_{cf}(\tau)]$ in first order of the 
spin coupling $J_{cf}$ and tracing out the 
conduction electron degrees of freedom, one obtains in the static limit 
($\omega=0$) the RKKY interaction operator between the local spins 
$\ve{S}_i$ and $\ve{S}_j$, 
\begin{eqnarray}\label{eqn:HRKKY1}
H^{RKKY}_{ij}= - J_{cf}^2 \left.\langle 
(\ve{S}_{i}\cdot \ve{s}_{i}) (\ve{S}_{j}\cdot \ve{s}_{j})
\rangle_{_c}\right|_{\omega=0} \ .
\end{eqnarray}
Here $\langle (\dots ) \rangle_{_c} := 
tr_c\{{\rm e}^{-\beta H} (\dots ) \}/Z_G$ denotes the thermal trace over 
the conduction electron Hilbert space. Using Wick's theorem, it can be 
written as,
\begin{eqnarray}\label{eqn:HRKKY2}
H^{RKKY}_{ij} \hspace*{-0.2cm}&=&\hspace*{-0.2cm} - \frac{J_{cf}^2}{4} \sum_{\alpha,\beta = x,y,z}\sum_{\sigma\sigma'}  
S_{i}^{\alpha}\, 
\sigma^{\alpha}_{\sigma\sigma'}\sigma^{\beta}_{\sigma'\sigma}\, 
S_{j}^{\beta}  \Pi_{ij}^{\sigma\sigma'}(0),\nonumber\\ &&
\end{eqnarray}
where $\Pi_{ij}^{\sigma\sigma'}$ is the conduction electron density propagator 
between the sites $i$ and $j$ as depicted diagrammatically 
in Fig.~\ref{fig:diagram_RKKY}. It has the general form,
\begin{eqnarray}
\Pi_{ij}^{\sigma\sigma'}(i\omega) = -\frac{1}{\beta}\sum_{\varepsilon_n}
G_{ji\,\sigma}(i\varepsilon_n+i\omega) G_{ij\,\sigma'}(i\varepsilon_n) \ . 
\end{eqnarray}
In the static limit it reads, 
\begin{eqnarray}
\Pi_{ij}^{\sigma\sigma'}(0) &=& - \int d\varepsilon\  f(\varepsilon)\ \times\\
&&\hspace*{-0.1cm}\left[
A_{ij\,\sigma}(\varepsilon) {\rm Re} G_{ij\,\sigma'}(\varepsilon) +
A_{ij\,\sigma'}(\varepsilon) {\rm Re} G_{ij\,\sigma}(\varepsilon) 
\right] \,,  \nonumber
\end{eqnarray}
where $A_{ij\,\sigma}(\varepsilon) = -{\rm Im}G_{ij\,\sigma}(\varepsilon +i0)/\pi$. 
Performing the spin contractions in Eq.~(\ref{eqn:HRKKY2}) and 
defining the longitudinal and the transverse polarization functions,
respectively, as
\begin{eqnarray}
\Pi_{ij}^{||}(0)   &=& \frac{1}{2}\sum_{\sigma} \Pi_{ij}^{\sigma\sigma}(0)\\
&=& - \sum_{\sigma}\int d\varepsilon\,  f(\varepsilon)\,
A_{ij\,\sigma}(\varepsilon) {\rm Re} G_{ij\,\sigma}(\varepsilon)\nonumber \\
\Pi_{ij}^{\perp}(0) &=& \frac{1}{2}\sum_{\sigma} \Pi_{ij}^{\sigma\,-\sigma}(0)\\ 
&=& - \sum_{\sigma}\int d\varepsilon\,  f(\varepsilon)\,
A_{ij\,\sigma}(\varepsilon) {\rm Re} G_{ij\,-\sigma}(\varepsilon) \ , \nonumber
\end{eqnarray}
one obtains the RKKY interaction Hamiltonian,  
\begin{eqnarray}\label{eqn:HRKKY3}
H^{RKKY} &=& \sum_{(i,j)} H^{RKKY}_{ij} \\
&=& - \sum_{(i,j)} \left[ K_{ij}^{||}\,  S_i^z S_j^z 
             - K_{ij}^{\perp}\, \left(S_i^xS_j^x + S_i^yS_j^y\right)\right] 
\nonumber
\end{eqnarray}
\vspace*{0.2cm}\noindent
where the sum runs over all (arbitrarily distant) pairs of localized 
spins $\ve{S}_i$ and $\ve{S}_j$, and
\begin{eqnarray}\label{eqn:KRKKY}
K_{ij}^{||}   =   \frac{1}{2} J_{cf}^2  \Pi_{ij}^{||}(0) \ , \qquad
K_{ij}^{\perp} = \frac{1}{2} J_{cf}^2 \Pi_{ij}^{\perp}(0) \ , 
\end{eqnarray}
are the longitudinal and transverse RKKY couplings,
respectively. As seen from Eqs.~(\ref{eqn:HRKKY3}) and (\ref{eqn:KRKKY}),
the RKKY interaction is in general anisotropic for a magnetized 
conduction band.

We now present explicitly the expressions for the special cases of a 
paramagnet and of a semimetal. For a paramagnetic conduction band we have
$G_{ij\,\sigma}=G_{ij, -\sigma}$ independent of spin. Hence, the RKKY coupling is
isotropic, and we have the paramagnetic RKKY Hamiltonian, 
\begin{eqnarray}\label{eqn:HRKKY_PM}
H^{RKKY}_{PM} = - \sum_{(i,j)} K_{ij}^{PM}\, \ve{S}_i\cdot
\ve{S}_j  \ ,
\end{eqnarray}
with 
\begin{eqnarray}
K_{ij}^{PM} = -\frac{J_{cf}^2}{2}\sum_{\sigma} 
\int d\varepsilon\, f(\varepsilon)\, 
A_{ij\,\sigma}(\varepsilon) {\rm Re} G_{ij\,\sigma}(\varepsilon) \ .
\phantom{xxx}
\label{eqn:JRKKY_PM}
\end{eqnarray}
For a semimetal, i.e., for a completely spin-magnetized conduction band
with majority spin $\sigma=\uparrow$ we have $A_{ij\,\downarrow}(\varepsilon)=0$, 
and the semimetallic RKKY Hamiltonian reads,
\begin{eqnarray}\label{eqn:HRKKY_FM}
H^{RKKY}_{FM} = - \sum_{(i,j)}&& \hspace*{-0.3cm} \left[
K_{ij}^{FM\ ||}\, S_i^zS_j^z \right. \\ 
&& \hspace*{-0.45cm}\left. + K_{ij}^{FM\ \perp}\, \left(S_i^xS_j^x+S_i^yS_j^y
\right)\right] \ , \nonumber
\end{eqnarray}
with 
\begin{eqnarray}
K_{ij}^{FM\ ||} &=& - \frac{J_{cf}^2}{2} 
\int d\varepsilon\, f(\varepsilon)\, 
A_{ij\,\uparrow}(\varepsilon) {\rm Re} G_{ij\,\uparrow}(\varepsilon) 
\label{eqn:JRKKY_FM1} \\ 
K_{ij}^{FM\ \perp} &=& - \frac{J_{cf}^2}{2} 
\int d\varepsilon\, f(\varepsilon)\, 
A_{ij\,\uparrow}(\varepsilon) {\rm Re} G_{ij\,\downarrow}(\varepsilon) \ . 
\phantom{xxxxx}
\label{eqn:JRKKY_FM2} 
\end{eqnarray}
The missing spin summation in Eqs.~(\ref{eqn:JRKKY_FM1}), (\ref{eqn:JRKKY_FM2}) 
as compared to Eq.~(\ref{eqn:JRKKY_PM}) indicates that 
in the completely magnetized band only the majority spin species contributes
to the coupling.
Note, however, that the transverse coupling $J_{FM\perp\ ij}^{RKKY}$ is still 
non-zero even in the ferromagnetically saturated case because of virtual
(off-shell) minority spin contributions represented by the real part, 
${\rm Re} G_{ij\,\downarrow}(\varepsilon)$ in Eq.~(\ref{eqn:JRKKY_FM2}).
The Curie temperature $T_C$, where the band magnetization vanishes, is 
controlled by the paramagnetic RKKY coupling, Eq.~(\ref{eqn:HRKKY_PM}),
while far below $T_C$ the carrier concentration $n_C$ is determined by the 
semimagnetic RKKY coupling, Eq.~(\ref{eqn:HRKKY_FM}). 

\section{NCA equations}\label{appendixB} 

The selfconsistent NCA equations for the pseudofermion ($f$) and 
slave boson ($b$) self-energies read ($\omega\equiv \omega+i0$)\\[-0.7cm]
\begin{eqnarray}
\Sigma_{f\sigma}(\omega) \hspace*{-0.15cm}&=&\hspace*{-0.15cm}
        V^2  \int d\varepsilon [1-f(\varepsilon)] 
        A_{c\sigma}(\varepsilon)  G_{b}(\omega - \varepsilon) 
	\label{eqn:nca_sigma_f} \\
\Sigma_b(\omega) \hspace*{-0.15cm}&=&\hspace*{-0.15cm}
	V^2 \sum_\sigma \int d\varepsilon f(\varepsilon) 
        A_{c\sigma}(\varepsilon)  G_{f\sigma}(\omega + \varepsilon) 
	\label{eqn:nca_sigma_b}\, , \\[-0.6cm] \nonumber
\end{eqnarray} 
with the auxiliary particle Green's functions,
$G_{f\sigma}(\omega)=1/\left[ \omega +\mu -\lambda -E_d 
                                -\Sigma_{f\sigma}(\omega) \right]$ and
$G_{b}(\omega)=1/\left[ \omega -\lambda  
                            -\Sigma_{b}(\omega) \right]$,
respectively. $\lambda$ is a positive parameter, taken to  
$\lambda\to\infty$ in order to effect the constraint on the 
auxiliary particle number operator, 
$\sum_{\sigma} \cre{f}{\sigma}\ann{f}{\sigma} + \cre{b}{}\ann{b}{}= 1$.
Note that these NCA equations are coupled to the equations
(\ref{eqn:tcb_bulk_green})--(\ref{eqn:avg_spin}) for the
interacting conduction electrons via the common chemical potential 
$\mu$ and via the conduction electron DOS of the interacting system
in presence of a dilute, but finite impurity concentration,
$A_{c\sigma}(\varepsilon)$. 
The Gd impurity electron Green's function is obtained from 
$G_{f\sigma}$, $G_{b}$ as,\\[-0.7cm]  
\begin{eqnarray}
G_{d\sigma}(\omega)
	&=&\int \frac{d \varepsilon} {e^{\beta\varepsilon}} 
 \bigl[A_{b}(\varepsilon)
  G_{f\sigma}(\varepsilon+\omega) - A_{f\sigma}(\varepsilon)
  G_{b}^*(\varepsilon-\omega) \bigr] \, . \nonumber\\ 
\label{eqn:nca_green_d}\\[-0.9cm] \nonumber
\end{eqnarray}
For an efficient and accurate method for numerically solving the set of 
equations (\ref{eqn:nca_sigma_f})--(\ref{eqn:nca_green_d}) see 
Ref.~[\onlinecite{Costi96}].
\bibliography{euobulk}

\begin{thebibliography}{42}
\bibitem[{\citenamefont{Note:}(.)}]{footnote1}
\bibinfo{author}{\bibnamefont{Note,}} \bibinfo{journal}{the reduction of the
  magnetic coupling observed in Ref.~[\onlinecite{Matsubara15}] only at high doping is
  indeed due to RKKY interaction, since the photodoping occurs not at the Fermi
  energy but into highly excited, short wavelength band states.}
\end{thebibliography}

\end{document}